# Checkerboard-type Zhang-Rice States in Overdoped Cuprate Superconductors

Xiongfang Liu [1,#], Kun Han [2,#], Yan Peng [3,#], Yuanjie Ning [1], Jing Wu [4], Zhaoyang Luo [5], Difan Zhou [1], Zhigang Zeng [1], Qian He [6], Chuanbing Cai [1], Mark. B. H. Breese [5,7], Ariando Ariando [5], Chi Sin Tang [7,*], George A. Sawatzky [8,*], Mi Jiang [3,9,*], Xinmao Yin [1,*]

[1] Shanghai Key Laboratory of High Temperature Superconductors, Center for Quantum Science and Technology, Department of Physics, Shanghai University, Shanghai 200444, China

[2] Information Materials and Intelligent Sensing Laboratory of Anhui Province, Institutes of Physical Science and Information Technology, Anhui University, Hefei 230601, China

[3] School of Physical Science and Technology, Soochow University, Suzhou 215006, China

[4] School of Electronic Science & Engineering, Southeast University, Nanjing, 211189, China

[5] Department of Physics, Faculty of Science, National University of Singapore, Singapore 117551, Republic of Singapore

[6] Department of Materials Science and Engineering, National University of Singapore, Singapore 117575, Republic of Singapore

[7] Singapore Synchrotron Light Source (SSLS), National University of Singapore, 5 Research Link, Singapore 117603, Republic of Singapore

[8] Department of Physics and Astronomy, University of British Columbia, Vancouver BC V6T 1Z1, Canada

[9] Jiangsu Key Laboratory of Frontier Material Physics and Devices, Soochow University, Suzhou 215006, China

[#]These authors contributed equally to this work.

[*]Corresponding author:

slscst@nus.edu.sg (C.S.T.); sawatzky@physics.ubc.ca (G.A.S);

jiangmi@suda.edu.cn (M.J.); yinxinmao@shu.edu.cn (X.Y.)

**Abstract**

Cuprate superconductors remain central to condensed matter physics due to their technological relevance and unconventional, incompletely understood electronic behavior. While the canonical phase diagram and low-energy models have been shaped largely by studies of underdoped and moderately doped cuprates, the overdoped regime has received comparatively limited attention. Here, we track the evolution of the electronic structure from optimal to heavy overdoping in $La_{2-x}Sr_xCuO_4$ (LSCO) using broadband optical spectroscopy across $x = 0.15$–$0.60$. The measured spectral changes—including the redistribution of Zhang–Rice–related spectral weight—are in qualitative agreement with determinant quantum Monte Carlo simulations of the three-orbital Emery model, which together indicate a pronounced reconstruction of the electronic structure beyond hole concentrations $x > 0.2$. Guided by these observations, we propose a spontaneous checkerboard-type Zhang–Rice electronic configuration that captures the coexistence of itinerant and localized carriers characteristic of the heavily overdoped state. Our results refine the doping-dependent Zhang–Rice–based framework for cuprates, illuminate how correlations persist deep into the overdoped regime, and provide new constraints on microscopic mechanisms of high-temperature superconductivity, with broader implications for correlated transition-metal oxides.

# 1. Introduction

The phase diagram of cuprate superconductors has been the focus of intense investigation and debate for nearly four decades since the discovery of high-temperature superconductivity (HTS) [1]. At low doping levels, the holes predominantly occupy the Zhang-Rice Singlet (ZRS) states, effectively lowering the Fermi level ($E_F$) to intersect the ZRS-derived band, leading to the emergence of finite conductivity. The ZRS picture was proposed as an effective low-energy building block enabling the reduction of a multi-band model to a single-band description or even a simpler $t$-$J$ model which discards all doubly occupied states [2-4]. Superconductivity sets in at moderate doping, characterized by an unconventional $d_{x^2-y^2}$-wave pairing symmetry on the Cu sites below the transition temperature ($T_c$) [5, 6]. As hole concentration increases further into the overdoped regime, the superconducting order weakens gradually and is suppressed eventually, with the system tending towards Fermi-liquid-like behavior [7, 8]. The enhancement of $T_c$ in the underdoped regime is generally attributed to the increased superfluid density [9]. In contrast, the heavily overdoped region lies well beyond the superconducting dome and remains relatively less explored, primarily due to synthetic challenges and the scarcity of high-quality single crystals [9, 10].

Early investigations of overdoped cuprate LSCO have primarily focused on the slightly overdoped regime (0.15<x≤0.3) with the aim of probing the electronic structure and the mechanisms responsible for HTS [10-13]. Spectroscopy measurements reveal a saturation behavior which points to a fundamental change in the electronic structure with further hole doping [5, 14]. In fact, the validity of the ZRS approximation is still a matter of ongoing debate, as a high-energy optical conductivity study reveals a strong mixture of singlet and triplet configurations in the lightly hole-doped Zn-LSCO single crystal [15]. Meanwhile, theoretical calculations have predicted the inapplicability of both the single-band Hubbard model and the ZRS framework in the overdoped regime [16]. Despite extensive studies in the slightly overdoped regime, a clear understanding of the precise evolution of the electronic structure and the validity of the ZRS model, particularly in the heavily overdoped regime (x>0.3), remains elusive.

The on-site Coulomb repulsion $U_{pp}$ between two electrons or two holes occupying the same ligand $p$ orbital is a key parameter in many strongly correlated systems. Although it has often been omitted in effective models (e.g., the $t$–$J$ model and the single-band Hubbard model), $U_{pp}$ can substantially shift ligand bands and influence both structural properties and magnetic interactions [17, 18]. Experimentally, the appearance, shift, or suppression of oxygen-derived spectral features reveals the interplay between $U_{pp}$ and the electronic structure [19]. In cuprates, $U_{pp}$ has been first estimated to be around 4 eV in earlier work [20], making its incorporation essential for understanding the evolution of the electronic structure and the validity of the ZRS picture in heavily overdoped LSCO.

In this work, we present a comprehensive spectroscopic investigation of LSCO across an extended overdoping range (0.15≤x≤0.60). Using high-resolution X-ray absorption spectroscopy (XAS) and optical spectroscopic ellipsometry (OSE) measurements, we observe the emergence of a new spectral feature concurrently in both the O *K*-edge XAS and optical conductivity ($\sigma_1$) spectra of heavily overdoped samples, which is absent in optimally and slightly overdoped LSCO. This newly identified feature points to a distinct modification of the LSCO electronic structure in which the quasiparticles can no longer be described in terms of ZRS or three-spin-polarons in the three-band model in the highly overdoped regime (Fig. 1a). In addition, our theoretical work within the three-orbital Emery model, using DQMC, indicates that the ZRS or three-spin-polarons description becomes inapplicable in the heavily overdoped regime and consistent with experimental observation of ZRS breakdown. Furthermore, our results predict a spontaneous formation of checkerboard-type ZR electronic configuration in the overdoped regime, offering new insight into the nature of superconductivity and its suppression in cuprate superconductors.

## 2. Materials preparation and quality

### 2.1. High crystalline quality of LSCO

Heavily overdoped LSCO (x > 0.3) bulk samples are challenging to synthesize due to the thermodynamic solubility limits and their structural instability [9]. Therefore, single-crystalline LSCO thin films with hole doping levels spanning 0.15 ≤ x ≤ 0.6 were synthesized on LaSrAlO$_4$ (LSAO) substrates using pulsed laser deposition (PLD) (Supplementary S1). All LSCO film samples have a thickness of ~90 nm.

Cross-sectional scanning transmission electron microscopy (STEM) high-angle annular dark-field (HAADF) images, shown in Fig. 1(b), reveal the high crystalline quality of the films. Similarly, uniform epitaxy is observed across the entire film with no observable structural defects. Figs. 1(c–d) and Figs. S1–S6 display the $2\theta$–$\theta$ scans and Reciprocal Space Mappings (RSM) from X-ray diffraction (XRD) measurements for all samples. The $2\theta$–$\theta$ scans demonstrate that all LSCO films exhibit a (00l)-preferred orientation and no additional peaks, indicating high phase purity and crystallinity. The RSM patterns for the $(004)_{HL}$, $(004)_{KL}$, $(-109)_{HL}$, and $(019)_{KL}$ reflections confirm the coherent lattice alignment between the LSAO substrate and the LSCO films, as well as the tetragonal symmetry of both materials (Supplementary S2) [21]. Furthermore, these observations suggest that both the in-plane and out-of-plane lattice parameters remain essentially unchanged over the wide doping range. Collectively, these characterizations establish the high crystalline quality of this LSCO series, providing a reliable platform for subsequent investigations of the optical response and electronic structure in strongly correlated cuprates.

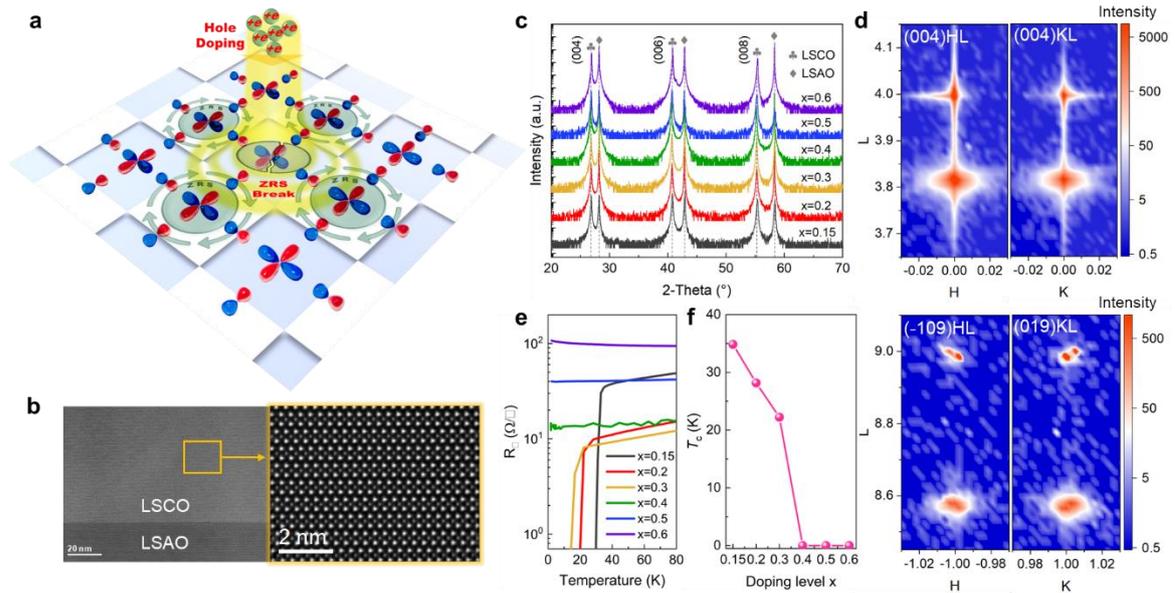

**Fig .1.** (a) Schematic diagram of the checkerboard-type Zhang–Rice electronic configuration within $CuO_2$ plane in heavily overdoped LSCO (b) Atomic resolution HAADF-STEM images of the representative LSCO (x=0.15)/LSAO interface along the (001) zone axis. (c) $2\theta$–$\theta$ scans for all LSCO (0.15≤x≤0.6)/LSAO via XRD. (d) RSM around (004) HL, (004) KL, (019) KL and (-109) HL for LSCO (x=0.15)/LSAO as characterized via XRD. (e) Resistance versus temperature for all LSCO (0.15≤x≤0.6)/LSAO samples and (f) the $T_c$ for LSCO/LSAO samples of different doping levels.

## 2.2. Transport measurement

The increase in hole concentration and the corresponding changes in $T_c$ are well-established characteristics of LSCO [22, 23]. To evaluate superconductivity across doping, the $T_c$ for all samples were determined from transport(resistivity) measurement (Figs.1(e-f)). The composition x=0.15 represents the optimally hole-doped LSCO with the highest $T_c$ of ~ 35 K [24]. As Sr (hole) doping increases, $T_c$ decreases ending with a rapid drop above x=0.3 to zero for x=0.4, in a manner consistent with the canonical LSCO superconducting dome. Consequently, superconductivity becomes completely suppressed in the heavily overdoped regime and does not reemerge within the explored composition range.

## 3. Optical analysis on electronic structure of LSCO

The superconductivity of LSCO evolves with hole doping, primarily due to holes entering the system causing the chemical potential to move into the top of the valence band of states resulting in (bad) metal behavior. To further investigate in the heavily overdoped regime, high-resolution XAS and OSE were employed to examine the unoccupied states and the optical response (dielectric function and optical conductivity) (Figs. 2a-b). By analyzing characteristic features in the optical spectra, we track the evolution of superconducting properties and electronic structure across hole concentrations.

### 3.1. XAS characterization

Figs. 2 and S7 display the XAS spectra at the O $K$-edge for both in-plane polarization ($\theta$=90 °) and out-of-plane polarization ($\theta$=25 °) at 300 K, measured in the total-electron-yield mode. Fig. 2(c) shows the in-plane O $K$-edge XAS spectra of LSCO for doping levels ranging from x = 0.15 to x = 0.6. These spectra exhibit doping dependence, intimately related to the O-2$p$–Cu-3$d$ hybridization. Two prominent features are observed, labeled $A$ (~528.6 eV) and $B$ (~530.1 eV). Feature $A$ corresponds to transitions from the O 2$p$ core level to the ZRS state, while feature $B$ corresponds to transitions from the O 2$p$ core level to the upper Hubbard band (UHB) [25, 26]. With the increase in Sr doping concentration, there is a corresponding increase in the intensity of feature $A$ and along with a redshift in its energy position. Meanwhile, the intensity of feature $B$ decreases, accompanied by a blueshift in photon energy. This behavior indicates spectral weight transfer from the UHB to the ZRS due to the incorporation of holes into the valence band, which is consistent with previous reports [25, 26]. These changes in intensity and energy positions of features $A$ and $B$ begin to saturate when x>0.2, while a prominent new shoulder feature labeled $A^*$ (~527.8 eV) emerges for x>0.2. The abrupt drop in XAS intensity observed at the doping concentration of x = 0.6 can be attributed to a spectral weight transfer to higher absorption energy (Supplementary S3).

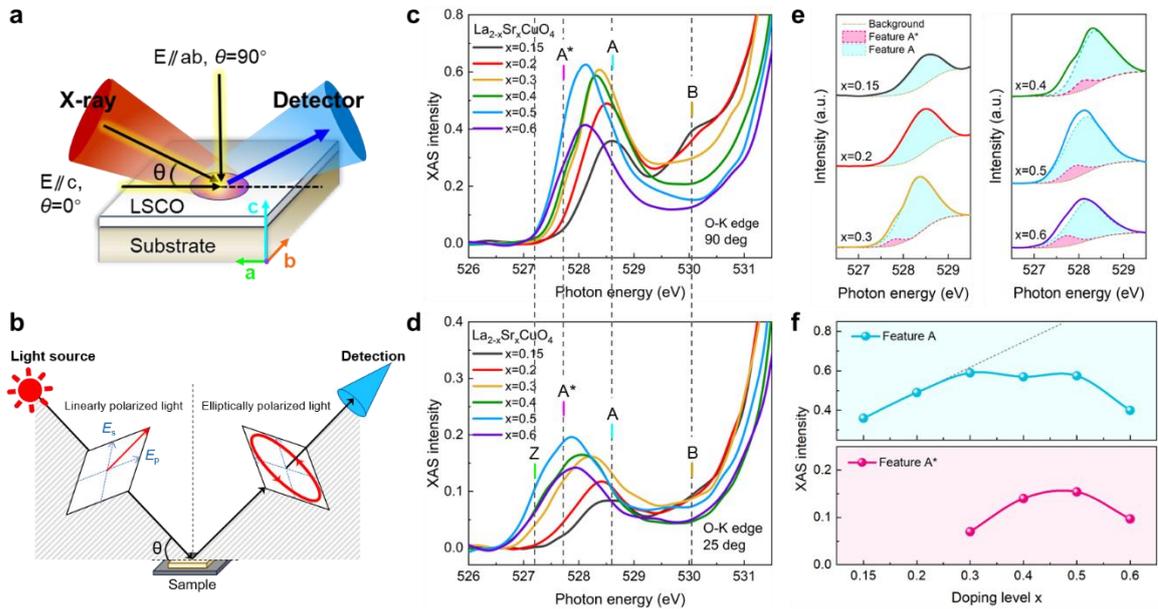

**Fig. 2.** (a) Schematic of XAS technique and its direction of photon polarization and incident angle. (b) Schematic of OSE experimental principle and methods in spectroscopic ellipsometry. (c) 90° (E//ab) and (D) 25° (E//c) incident XAS at the O $K$-edge for LSCO/LSAO samples of different doping concentration. (E) The fitting results of (C) for identify the feature $A^*$ marked pink filling and the feature $A$ marked light blue filling. (F) The intensity of 90° XAS spectra of respective feature $A^*$ and $A$ at O $K$-edge.

For comparison, the out-of-plane polarization O $K$-edge XAS spectra of LSCO for various doping levels are shown in Fig. 2(d). The out-of-plane polarization spectra inevitably contain a finite contribution from the in-plane polarization response (Supplementary S3). The spectral weight of the observed features is significantly reduced as the polarization becomes perpendicular to the $CuO_2$ plane but exhibits a similar dependence on doping concentration [12]. The out-of-plane XAS spectra also display signals associated with the ZRS (feature $A$), UHB (feature $B$) and the newly identified feature $A^*$. In addition, there is the appearance of another feature labelled $Z$, related to O $2p_z$ character, unique to measurements at this polarization geometry and has been reported in an earlier study [27]. Interestingly, the out-of-plane XAS also registers feature $A^*$, located at ~527.7 eV, at doping level $x>0.2$, similar to the in-plane XAS measurements. Therefore, the observation of feature $A^*$ represents a new and intriguing finding in the O $K$-edge XAS spectra.

A spectral fitting method was applied to the in-plane O $K$-edge XAS spectra to identify feature $A^*$, as shown in Fig. 2(e). According to the fitting results, feature $A^*$ emerges adjacent to feature A for doping level x>0.2, whereas it is absent when x≤0.2. As the doping level increases beyond 0.2, feature $A^*$ becomes progressively more pronounced. The appearance of feature $A^*$ at x>0.2 suggests that excessive hole doping in LSCO modifies the electronic structure, resulting in the formation of a new unoccupied state above $E_F$. The doping dependence of the spectral weight of features $A^*$ and $A$ from fitting results is illustrated in Fig. 2(f). Feature $A$ exhibits an approximately linear doping dependence for x≤0.2. However, in the heavily overdoped regime (x>0.2), Feature $A$ deviates from this linear trend, with the spectral weight decreasing in magnitude and gradually approaching saturation, indicating a weaker doping dependence. The saturation behavior of feature $A$ for x>0.2 is consistent with previous descriptions of XAS spectra [5, 28]. The development of feature $A^*$ after the saturation of feature $A$ suggests a transformation in the nature of ZRS-related electronic states in the heavily overdoped regime.

## 3.2. OSE Characterization

Detailed optical characterization was performed on all LSCO samples using optical spectroscopic ellipsometry (OSE) at 300 K over the photon energy range of 0.5-3.5 eV (Supplementary S4). The real and imaginary components of the dielectric function, $\varepsilon(\omega) = \varepsilon_1(\omega) + i\varepsilon_2(\omega)$, can be independently obtained by OSE without requiring a *Kramers–Kronig* transformation [29]. The real component of the dielectric function, $\varepsilon_1$, reflects the polarizability of the material, and its spectrum is shown in Fig. 3(a). The zero-crossing point in the $\varepsilon_1$ spectrum indicates the presence of high-mobility carriers and is associated with conventional metallic plasmon excitations [30]. The zero-crossing energy of $\varepsilon_1$ increases for $0<x\leq0.2$, but decreases for $x>0.2$ (Fig. 3(b)). The downward shift of $\varepsilon_1$'s zero-crossing indicates an unexpected reduction in the effective carrier density in heavily overdoped LSCO in doping concentration $x>0.2$. The imaginary part, $\varepsilon_2$, can be used to derive the optical conductivity spectrum, $\sigma_1(\omega)=\omega\varepsilon_2(\omega)/4\pi$, which provides insights into how the electrical conductivity of LSCO evolves with increasing hole doping.

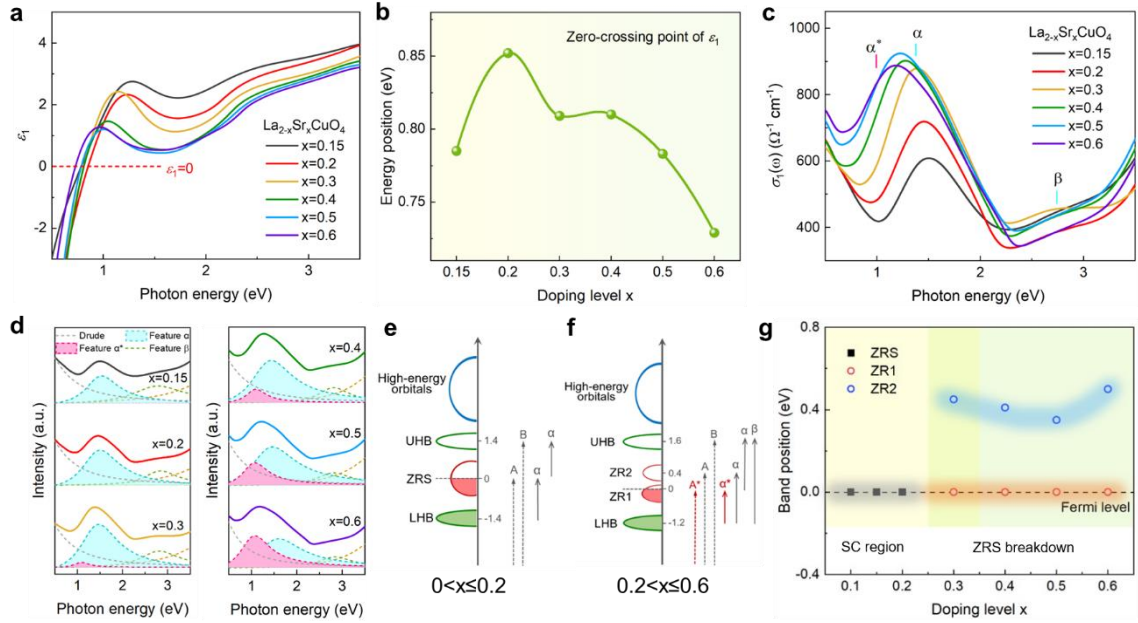

**Fig. 3** (a) The real components of the dielectric function $\varepsilon_1$, (b) $\varepsilon_1$ zero-crossing position and, (c) optical conductivity $\sigma_1$ spectra of LSCO/LSAO samples at varying doping concentration $x$. (d) represents the Drude-Lorentz fit, which is decomposed into a Drude component and several Lorentz components, $\alpha^*$ and $\alpha$, denoted by the pink filled and light bule filled regions, respectively. Estimated electronic structure for (e) LSCO (0<$x$≤0.2) and (f) LSCO (0.2<$x$≤0.6). The arrows and the respective letters denote the interband transitions in different doped LSCO as registered by the XAS and OSE measurements previously displayed in Figs. 2(c) and 3(c), respectively. (g) Schematic displaying the ZRS breakdown consist of clear electronic structure at $E_F$ in heavily overdoped LSCO.

The $\sigma_1$ spectrum in the energy range of 0.5–3.5 eV shows relatively low spectral weight at x≤0.2. Subsequently, the spectral weight increases rapidly for x>0.2 [Fig. 3(c)]. The Drude–Lorentz model was employed to fit the $\sigma_1$ spectra of all LSCO samples, enabling precise determination of the energy positions and spectral weights of the individual features [31]. The corresponding fitting curves are presented in Fig. 3(d). For x≤0.2, the $\sigma_1$ spectrum comprises a Drude term plus three Lorentz features (four components in total). The Drude response at $\omega$=0 corresponds to the zero-crossing point in the $\varepsilon_1$ spectrum and reflects the metallic properties associated with delocalized carriers [32]. Feature $\alpha$ (~1.5 eV), $\beta$ (~2.8 eV) and a higher-energy feature (>3.2 eV) appear consistently in all LSCO samples and originate from interband electronic transitions— from the ZRS to the UHB, from the lower Hubbard band (LHB) to the UHB, and from higher-energy excitations, respectively [33-35].

Notably, a new feature $\alpha^*$ (~1.1 eV) gradually emerges for x>0.2 in the $\sigma_1$ spectrum, highlighted in pink in Fig. 3(d). Feature $\alpha^*$ is absent in optimally doped (x=0.15) and slightly overdoped (x=0.2) LSCO samples but appears in heavily overdoped LSCO (x>0.2), with its intensity increasing monotonically as the doping level rises. Simultaneously, the spectral weight of feature $\alpha$ transfers progressively to feature $\alpha^*$ with increasing doping, reflecting changes in the ZRS-related electronic structure.

### 3.3 New electronic structure in heavily overdoped LSCO

Combined with a systematic analysis of XAS and $\sigma_1$ spectra, the concurrent detection of the new feature $A^*$ in the O $K$-edge XAS spectrum and the new feature $\alpha^*$ in the $\sigma_1$ spectrum is an interesting finding. The emergence of both features signals significant changes in the electronic structure of LSCO as the doping level varies. Information extracted from the features observed in the O $K$-edge XAS and $\sigma_1$ spectra provides insight into the electronic structure evolution, and the corresponding energy band diagrams for 0<x≤0.2 and x>0.2 in LSCO are presented in Figs. 3(e-f).

For LSCO samples with 0<x≤0.2, the ZRS state exists as a half-filled band at $E_F$, while the UHB is unoccupied and the LHB is fully occupied [Figs. 3(e)]. The energy separation between the UHB and LHB corresponds to the Hubbard Coulomb potential. Within this doping range, holes enter the ZRS, transforming the Mott insulating parent compound LCO into superconducting LSCO, with the highest $T_c$ achieved at optimal doping around x=0.15. Upon further hole doping beyond this level, $T_c$ begins to decrease, while the ZRS remains relatively stable in the range 0<x≤0.2. This scenario is consistent with previous studies [12, 25, 35].

However, as the hole doping concentration continues to increase, the ZRS spectral weight begins to saturate and overlap, leading to significant interactions and interference at high hole densities. This enhanced hole density on the oxygen sites amplifies the additional Coulomb repulsion $U_{pp}$ on the oxygen sites when two holes occupy orbitals on the same oxygen atom [5]. The heavily overdoped system becomes less stable energetically as $U_{pp}$ grows, destabilizing the ZRS manifold. Once the ZRS manifold is exhausted, new electronic states emerge. Consequently, the ZRS begins to breakdown, giving rise to two new bands, here labeled "Zhang-Rice feature 1 (ZR1)" and "Zhang-Rice feature 2 (ZR2)". The ZR1 band remains near $E_F$, while the ZR2 band shifts to higher energies above $E_F$ [Fig. 3(f)]. The transformation of the originally broad ZRS feature into a narrower ZR1 band near $E_F$ implies increased charge localization, leading to a concomitant reduction in the effective carrier density in the heavily overdoped regime, as shown in Fig. 3(b). In addition, the emergence of two distinct electronic states, ZR1 and ZR2, respectively, supports an alternating Zhang–Rice motif in the electronic landscape. The new feature $A^*$ observed in the XAS can be attributed to transitions into the partially unoccupied ZR1 band, whereas feature $\alpha^*$ detected in the $\sigma_1$ spectrum originates from electronic transitions between the LHB and the half-filled

ZR1 band. The evolution of the band positions near $E_F$ as a function of doping is illustrated in Fig. 3(g). This diagram clearly demonstrates the breakdown of the ZRS and its splitting into two distinct components in heavily overdoped LSCO.

## 4. Simulating three-orbital Emery model with determinant quantum Monte Carlo (DQMC)

To further investigate the origin of the anomalous low-energy behavior, we perform numerically unbiased determinant quantum Monte Carlo (DQMC) simulations within the three-orbital Emery model (Supplementary S7) [36]. As illustrated in Fig. 4(a), the three-orbital model includes only $d_{x^2-y^2}$ and $p_{x,y}$ orbitals, with the hole-language hopping phase conventions indicated. Without assuming the existence of ZRS in advance, the three-orbital model serves as a minimal model that captures the essential physics without imposing a priori ZRS [37]. We use a canonical parameter set with the $d$-$p$ hopping $t_{pd}$ taken as the energy unit. The Hamiltonian includes on-site Coulomb repulsion $U_{dd}$ on the Cu $d$-orbitals and $U_{pp}$ on the O $p$-orbitals respectively, and reads as follows (Supplementary S7):

$$\widehat{H} = \widehat{E}^s + \widehat{K} + \widehat{U}$$
$$\widehat{E}^s = (\epsilon_d - \mu)\sum_{i\sigma}\hat{n}^d_{i\sigma} + (\epsilon_p - \mu)\sum_{j\sigma}\hat{n}^p_{j\sigma}$$
$$\widehat{K} = \sum_{\langle ij\rangle\sigma}t^{ij}_{pd}(\hat{d}^\dagger_{i\sigma}\hat{p}_{j\sigma} + h.c.) + \sum_{\langle jj'\rangle\sigma}t^{jj'}_{pp}(\hat{p}^\dagger_{j\sigma}\hat{p}_{j'\sigma} + h.c.)$$
$$\widehat{U} = U_{dd}\sum_i\hat{n}_{i\uparrow}\hat{n}_{i\downarrow} + U_{pp}\sum_j\hat{n}_{j\uparrow}\hat{n}_{j\downarrow}$$

Fig. 4(b) shows the local density of states (DOS), extracted via the maximum entropy analytic continuation (MaxEnt) [38, 39], over a substantial hole-doping range $p_h$ from 0.1 to 0.6. As the doping level, $p_h$, increases, the red Cu-derived peak above $E_F$, commonly associated with the UHB, moves toward higher energy with spectral weight significantly reduced [40]. The broad feature centered near ~3 eV is the charge-transfer band located below $E_F$. Due to a relatively high simulation temperature, the LHB is largely broadened. In the simulated hole doping range of $p_h \leq 0.4$, the ZRS remains intact and shows strong $p$-$d$ hybridization. Nonetheless, as the doping level reaches $p_h$ =0.5, the spectra show split features, supporting the optical observation of ZRS breakdown. Since the ZRS is formed by the linear combination of one hole on Cu 3$d$ orbital and the other on the four ligand O-2$p$ orbitals, it is sensitive to the hole densities on Cu and O [3]. In the charge transfer regime, the doped holes mainly reside on O orbitals due to the higher repulsion on Cu orbitals. This may lead to a gradual deviation of Cu-O hole concentration and thereby strongly affecting the stability of the ZRS [3]. As a result, the Cu-O coupling strength in the UHB and the ZRS is substantially weakened, especially when $p_h$ exceeds 0.4. More detailed investigation of the $\boldsymbol{k}$-resolved spectral dispersion has been conducted in a recent study [41].

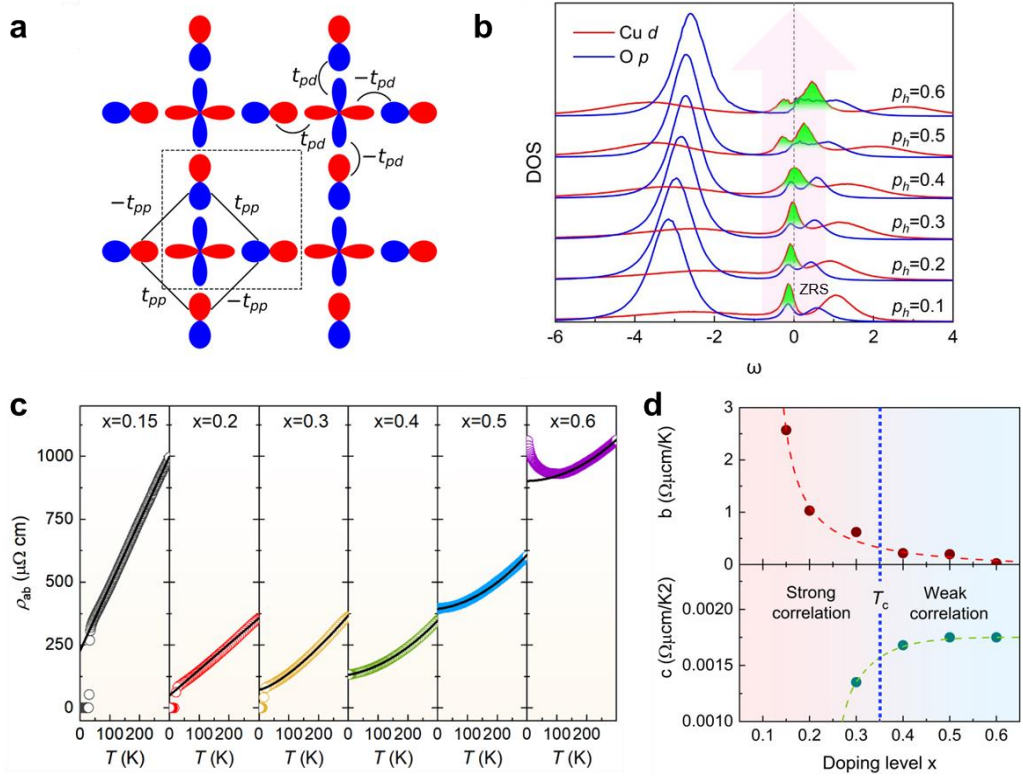

**Fig. 4** (a) A schematic illustration of Cu and O orbitals in three-orbital Emery model. the superscripts of $t_{pd}$ and $t_{pp}$ represent the sign of hopping phase. Component with Different colors represent opposite phase factor. The unit cell is included by the dashed square. (b) Simulated DOS for LSCO with varying doping levels n. The Red (blue) solid line represents Cu (O) orbitals. (c) The Fitting results of the in-plane resistivity of LSCO with varying doping levels, whereas the colorful scatters are corresponding to $\rho(T)$ values by transport measurements and the black solid lines represent fits to the data below 300 K using the expression $\rho(T)=a+bT+cT^2$. (d) Doping evolution of the coefficients b (T-linear resistivity component) and c ($T^2$ resistivity component) of expression $\rho(T)=a+bT+cT^2$.

## 5. Discussion

### 5.1 Anomalous Electrical Resistivity of heavily overdoped LSCO

According to the established LSCO phase diagram, the system is commonly understood to evolve from strange-metal to Fermi-liquid behavior as the doping level increases from x=0.15 to 0.60 [8]. Therefore, the resistivity behavior of LSCO change significantly across different doping levels. The expression $\rho(T)=a+bT+cT^2$ is an appropriate model for describing the resistivity of overdoped LSCO samples (Supplementary S6). The parallel-resistivity fits (solid black curves in Fig. 4c) are nearly indistinguishable from the experimental $\rho(T)$ curves over the range 0.15≤x≤0.6. Resistivity decreases with increasing x up to 0.4, then rises at higher overdoping levels, consistent with previous studies on LSCO [42, 43]. A distinct linear $\rho-T$ relationship is observed down to $T_c$ in LSCO (0.15≤x≤0.2) samples, following the simplified relation

$\rho(T) = a+bT$, with c=0. The persistence of this strange-metal behavior up to 300 K likely reflects the presence of strong electronic correlations. Nevertheless, at higher doping levels (x>0.2), the $\rho(T)$ curves exhibit pronounced nonlinear behavior that is well described by the full expression $\rho(T)=a+bT+cT^2$.

The doping evolution of the coefficients b and c obtained from corresponding $\rho-T$ fits is shown in Fig. 4d. The coefficients b, associated with strong electron-electron scattering, is observed to decrease rapidly with increasing x. Notably, b tends toward low value as $T_c$ disappears, suggesting that strong electronic correlations are intimately linked to superconductivity [7]. Conversely, as the coefficient $b$ decreases and $c$ grows in the fit $\rho(T) = a+bT+cT^2$, the system exhibits a gradual crossover toward Fermi-liquid–like behavior that extends deep into the heavily overdoped regime. Nevertheless, the resistivity deviates from the canonical Fermi-liquid form $\rho(T) \propto T^2$ due to the coexistence of linear and quadratic temperature components. At the highest dopings ($x > 0.5$), LSCO may even approach a metal–insulator transition. Previous studies of bosonic excitations in cuprates have also demonstrated similar $\rho-T$ relationships and doping dependences [8, 44].

## 5.2 ZRS breakdown and electronic correlations

Taken together with our new findings on the electronic structure of heavily overdoped LSCO (Figs. 3e–f), the simple picture of "one additional ZRS hole per Sr atom" no longer strictly applies. Instead, the extra holes modify the electronic structure in a fundamentally different way, leading to the breakdown of the ZRS manifold. The crossover to a nonlinear resistivity regime appears to occur concomitantly with the breakdown of the ZRS. The heavily overdoped regime therefore hosts distinct many-body phenomena. With increasing $x$, resistivity becomes increasingly nonlinear while the superconducting $T_c$ decreases—suggesting a progressive weakening of strong electronic correlations accompanying the ZRS breakdown [34].

The breakdown of the ZRS occurs concurrently with the reduction of strong correlations in overdoped cuprates. This may suggest that the carriers participating in superconductivity become less effective at engaging in spin fluctuations and Cooper pairing effectively reducing both the superfluid density and the pairing strength [10, 45-47]. Consequently, $T_c$ diminishes and eventually vanishes. This weakened-correlation scenario drives heavily overdoped LSCO toward a state characterized by nonlinear resistivity.

## 6. Conclusion

Our combined XAS and OSE measurements reveal significant electronic structure changes in heavily overdoped LSCO. This is evidenced by the appearance of distinct spectral features at doping levels x > 0.2, signaling the saturation and destabilization of the ZRS picture. This ZRS breakdown is accompanied by the formation of new, reduced-correlation bands, reflected in both spectroscopic signatures and coinciding with a doping-induced crossover in resistivity from linear to nonlinear behavior. These observations indicate that the heavily overdoped regime cannot be described solely by the traditional strong-correlation ZRS framework and instead evolves toward a moderate-correlation state.

Further analysis suggests that the originally broad ZRS feature progressively evolves into two narrower subbands, denoted here as ZR1 and ZR2. The ZR1 band, located near the $E_F$, retains the original ZRS-derived character, whereas the ZR2 band is more localized and exhibits a diminished ZRS contribution. This ZRS breakdown reflects an increasing differentiation of the electronic states in the heavily overdoped regime. Consistent with this, the observation of a dual-component electrical resistivity indicates the coexistence of two distinct electronic subsystems within the same $CuO_2$ plane. Such coexistence implies that heavily overdoped cuprates may favor an inhomogeneous correlated state and cannot be regarded as conventional Fermi liquid [7].

In heavily overdoped regime of approximately 0.5 hole per O square, one can imagine a spontaneous configuration in which half of the Cu sites host a ZRS ($3d^{9+}$ hole, S = 0) while the remaining half retain the $Cu^{2+}$-like ($3d^9$, S=1/2) character. These two types of Cu sites may arrange alternatively to form a checkerboard-type electronic pattern, as illustrated in Fig. 1a. In such a configuration, one sublattice consists of O squares each hosting a hole that forms a ZRS with the central Cu of $dx^2-y^2$ symmetry, while the other sublattice comprises $Cu^{2+}$-like sites without additional holes. At precisely 0.5 hole doping, this alternating arrangement would likely yield an insulating system (probably related to the transport results of LSCO(x=0.6) in Fig. 4c). With further hole doping, the additional holes are expected to occupy oxygen orbitals with s-like $dx^2+y^2$ symmetry around the $Cu^{2+}$-like sites and hybridize with the $3dz^2-r^2$ orbital, leading to both in-plane and out-of-plane polarization features at the O absorption edge. However, XAS cannot distinguish between $x^2+y^2$ and $x^2-y^2$ orbital combinations.

Therefore, hole-doped cuprates may exhibit two characteristic regimes of hole incorporation. At low doping, holes propagate predominantly as ZRS or three-spin polaron states; near 0.5 hole per O square, the system transits to a checkerboard-type configuration with alternating ZRS and $Cu^{2+}$-like sites. In principle, intermediate phases with varying proportions of ZRS and $Cu^{2+}$-like character could exist between these limits. This hypothetical alternating configuration finds an intriguing parallel in the nickelates, where recent resonant X-ray scattering measurements on $NdNiO_3$ have revealed alternating Ni sites with large spin moments and strong antiferromagnetic Ni–O coupling [48]. Such a bond-disproportionated state, characterized by alternating Ni–O ZRS–like and $Ni^{2+}$-like sites, provides a compelling analogue for understanding how heavily overdoped cuprates might undergo electronic phase separation or bond disproportionation at extreme doping levels.

The overdoped regime of the cuprate phase diagram, once regarded as a conventional

Fermi liquid limit, is increasingly recognized as a key frontier for understanding correlated electron behavior. At high doping, the system undergoes a dramatic reconstruction of the electronic structure while retaining vestigial correlation effects. Meanwhile, overdoped cuprates exhibit a subtle interplay between itinerant and localized carriers, suggesting proximity to electronic phase separation states.

Importantly, the overdoped regime offers a unique opportunity to probe the intrinsic mechanism of superconductivity under conditions where competing pseudogap and antiferromagnetism phenomena are largely suppressed. By studying how the pairing strength, coherence, and carrier dynamics evolve across the overdoped regime, one may isolate the essential electronic interactions responsible for superconducting condensation. Looking forward, systematic investigations combining high-resolution spectroscopies (RIXS, ARPES, and ultrafast optical probes) with quantum many-body calculations will be crucial to elucidate how the checkerboard-type electronic pattern evolves beyond the optimal doping level. Ultimately, the heavily overdoped cuprates provide a clean, tunable platform for unraveling the universal pairing mechanism in high-temperature superconductors.

**Acknowledgments**
We appreciate Yuan Li (Institute of Physics, Chinese Academy of Sciences) for valuable discussion. We acknowledge support by National Natural Science Foundation of China (Grant Nos. 12374378, 52172271, 52307026, 52477022), the National Key R&D Program of China (Grant No. 2022YFE03150200), Shanghai Science and Technology Innovation Program (Grant No. 23511101600). This work is supported by the Ministry of Education (MOE), Singapore, under its Tier-2 Academic Research Fund (AcRF) (Grant No. MOE-T2EP50124-0003). Chi Sin Tang acknowledges the support from the NUS Emerging Scientist Fellowship. Yan Peng and Mi Jiang acknowledge the support from National Natural Science Foundation of China (Grant No. 12174278). George A. Sawatzky is funded by the Quantum Matter Institute (QMI) at University of British Columbia and the Natural Sciences and Engineering Research Council of Canada (NSERC). The authors would like to acknowledge the Singapore Synchrotron Light Source for providing the facility necessary for conducting the research. The Laboratory is a National Research Infrastructure under the National Research Foundation, Singapore. Any opinions, findings, and conclusions or recommendations expressed in this material are those of the author(s) and do not reflect the views of National Research Foundation, Singapore.

**Author contributions**
X.L., K.H., Y.P. contributed equally to this work. X.Y. conceived the project. K.H., Z.Z. synthesized the samples and X.L., J.W., D.Z. performed the XRD experiments and analyzed the data. Z.L., Q.H., A.A. performed the STEM experiments and analyzed the data. X.L., C.S.T., M.B.H.B., G.A.S. performed the XAS experiments and analyzed the data. X.L., Y.N., X.Y. performed the SE experiments and analyzed the data. Y.P., M.J. performed the theoretical calculations and analyzed the result. X.L., Y.P., C.S.T. wrote the manuscript, with input from all the authors.

**Conflict of Interest**
The authors have no conflicts to disclose.

**Data availability**
The data that support the findings of this study are available from the corresponding authors upon reasonable request.